\documentclass{ws-ijmpa}

\begin{document}

\title{
RESULTS FROM THE ANALYSIS OF CRYSTAL BALL\\
MESON PRODUCTION MEASUREMENTS AT BNL
}
\author{
\large{R.A. Arndt, W.J. Briscoe, I.I. Strakovsky, and R.L.
Workman}}
\address{
Center for Nuclear Studies, Department of Physics \\
The George Washington University, Washington, D.C. 20052 \\
}

\maketitle

%\pub{Received (23 July 2006)}

\begin{abstract}
The Crystal Ball spectrometer, with its nearly complete angular
coverage, is an efficient detector of photon and neutron final
states.  While installed in the C6 beamline of the Alternating
Gradient Synchrotron (AGS) of Brookhaven National Laboratory
(BNL), this feature was used in a series of precise measurements
of reactions with all-neutral final states.  Here we concentrate
on the analysis of data from the pion-induced reactions:
$\pi^-p\to\gamma n$, $\pi^0n$, $\eta n$, and $\pi^0\pi^0 n$.

\end{abstract}

%%%%%%%%%%%%%%%%%%%%%%%%%%%%%%%%%%%%%%%%%%%%%%%%%%
\section{Overview}

\begin{figure}[htbp]
\centerline{
      \includegraphics[height=0.6\textwidth, angle=-90]{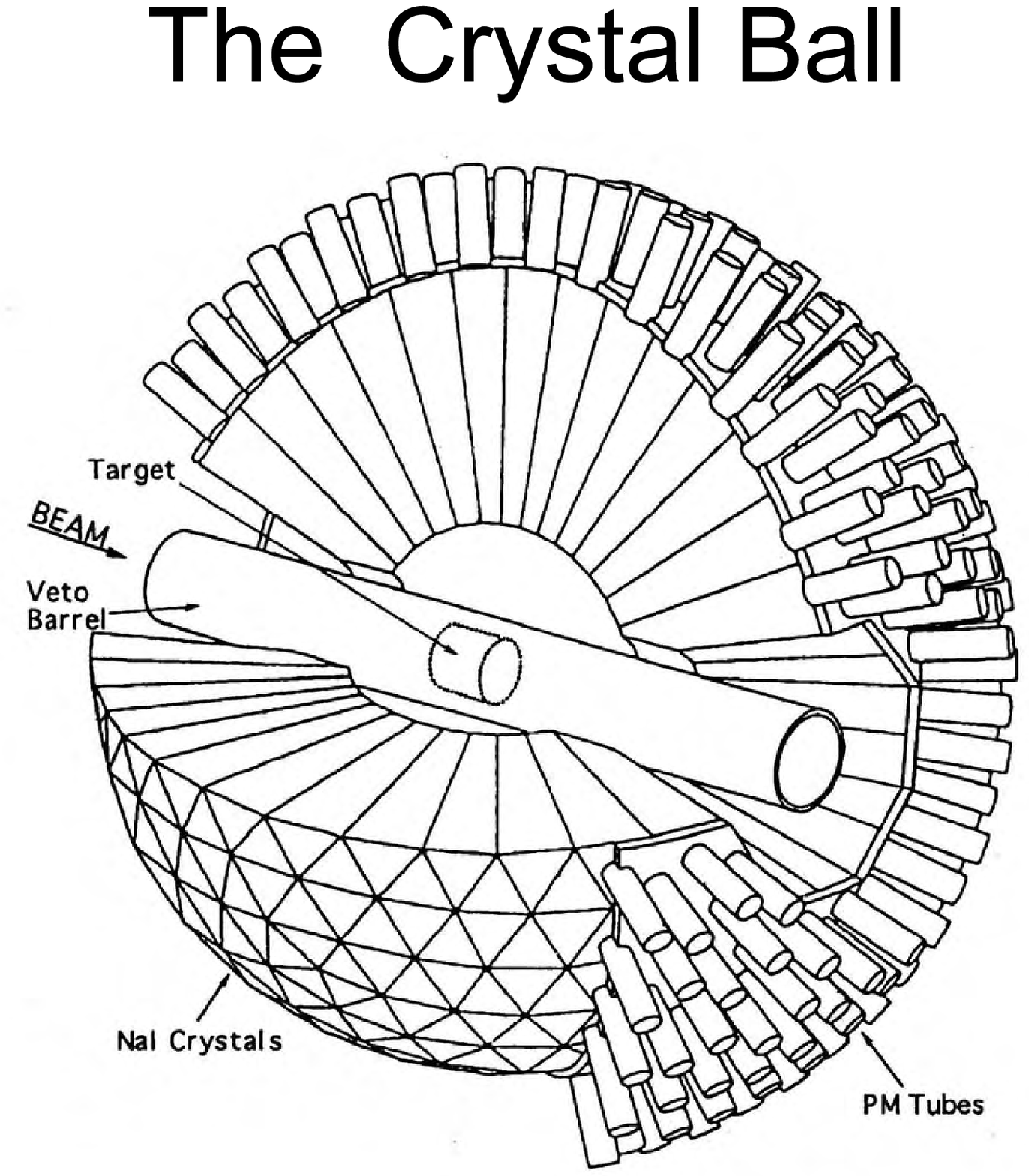}
}
      \caption{Crystal Ball detector.} \label{fig:g1}

\end{figure}

The Crystal Ball detector (Fig.~\ref{fig:g1}), with its nearly
4$\pi$ angular coverage (94\% of 4$\pi$, utilizing 672 NaI
crystals) is capable of measuring neutral final-state particles
with high efficiency. The E913/E914 measurements reported here
have provided very precise cross sections with broad angular
coverage. Such data are invaluable for detailed partial-wave
analyses. Below we consider how these data have impacted ongoing
studies of the low-lying baryon resonances.
%%%%%%%%%%%%%%%%%%%%%%%%%%%%%%%%%%%%%%%%%%%%%%%%%%%%%%%%
%%%%%%%%%%%%%%%%%%%%%%%%%%%%%%%%%%%%%%%%%%%%%%%%%%%%%%%%%

%%%%%%%%%%%%%%%%%%%%%%%%%%%%%%%%%%%%%%%%%%%%%%%%%%%%
\section{Charge-Exchange Reaction}

Measurements of the charge-exchange reaction were made from
threshold to T$_{\pi^-}$ = 625~MeV. Two regions were analyzed
separately: the 64 to 212~MeV T$_{\pi^-}$ range, dominated by
the $\Delta(1232)$ resonance~\cite{cex1}, and the 524 to
625~MeV range, which is clearly influenced by the opening
$\eta n$ channel~\cite{cex2}.

\begin{figure}[htbp]
      \includegraphics[height=0.5\textwidth, angle=90]{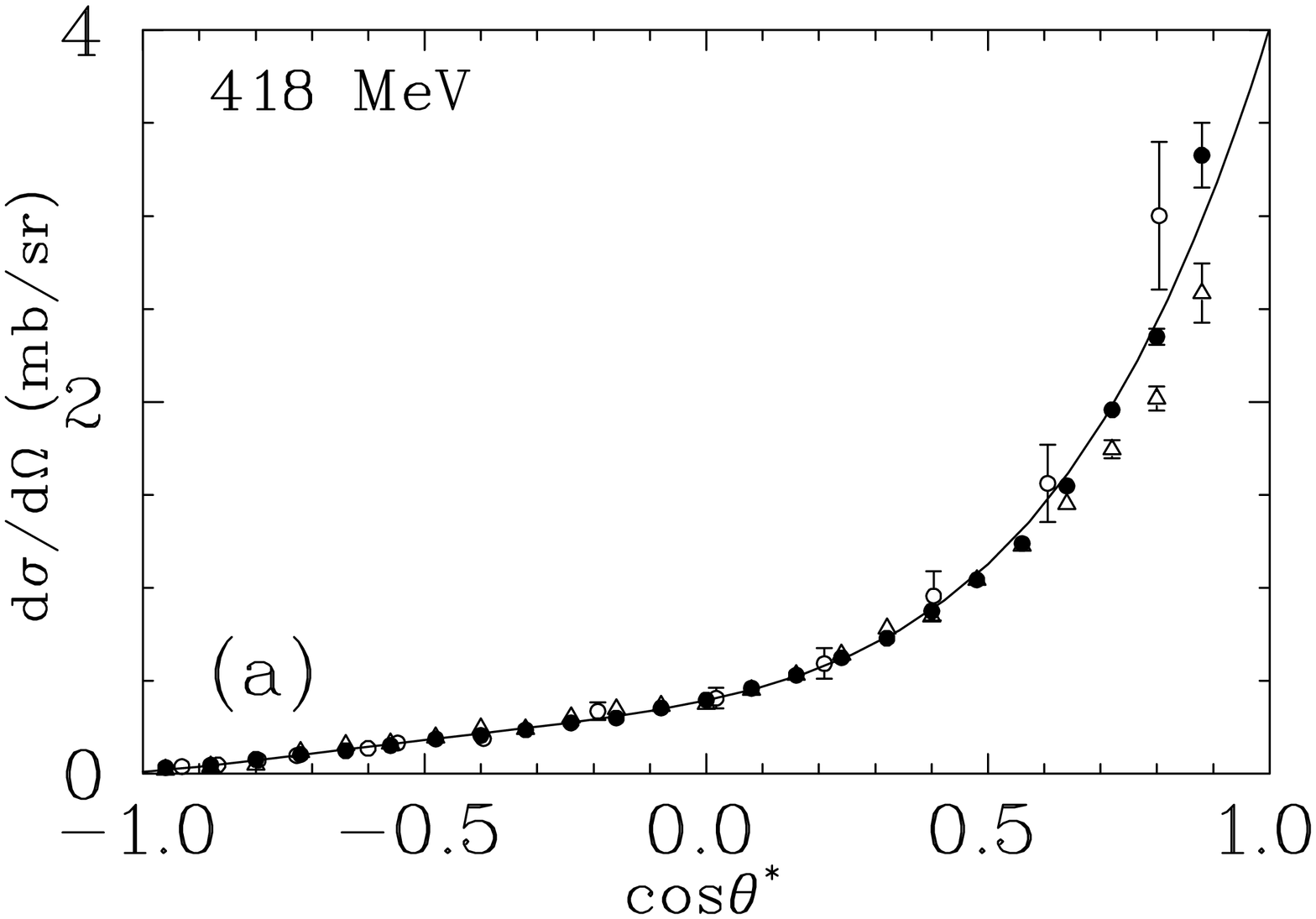}\hfill
      \includegraphics[height=0.5\textwidth, angle=90]{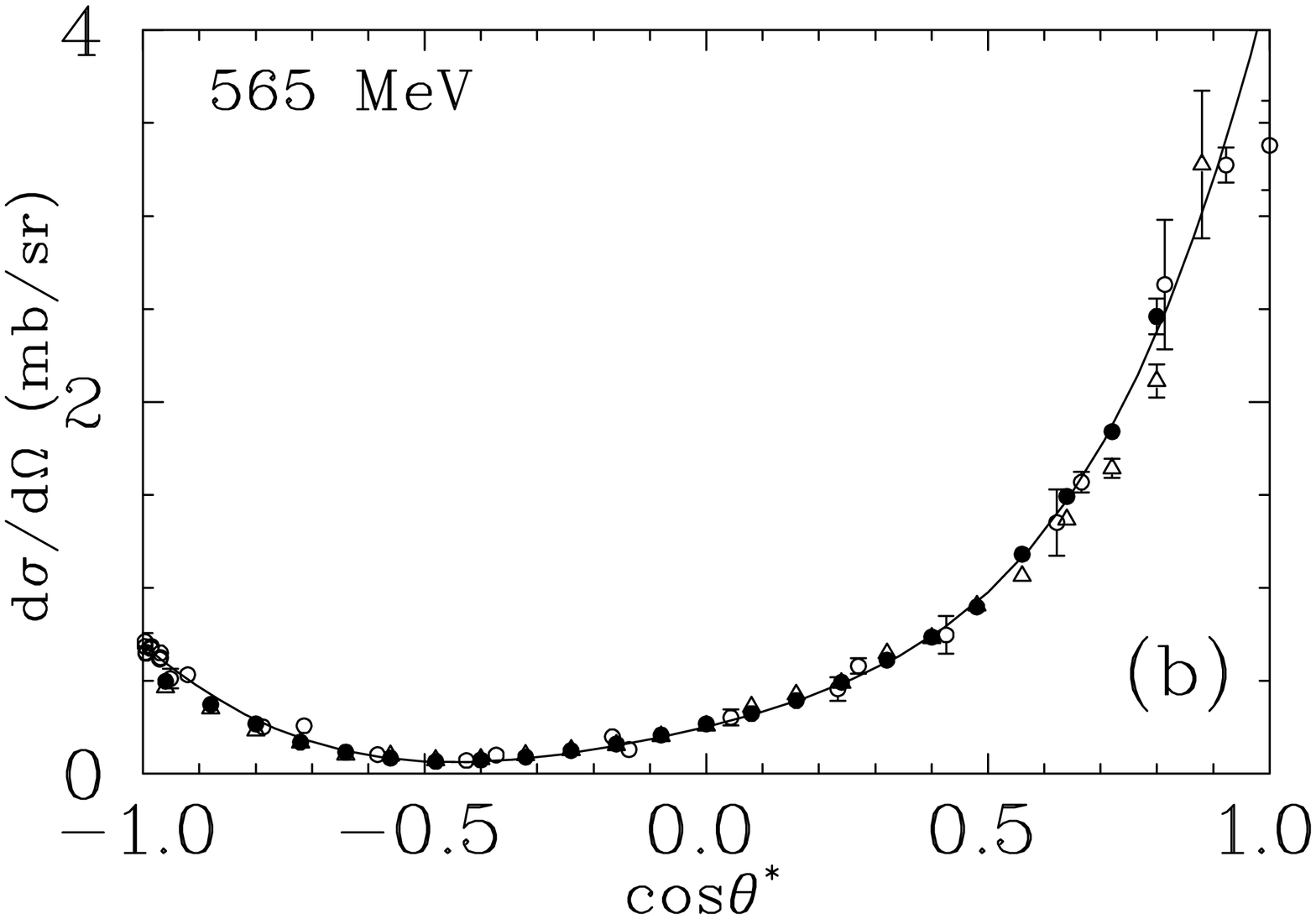}
      \caption{Differential cross sections for
       $\pi^-p\to\pi^0n$.  The $LH_2$ data (filled
       circles) are normalized to the central part
       of the $CH_2$ spectra (open triangles).  Solid
       lines show the FA02 predictions~\protect\cite{fa02}.
       E913/E914 data are from Ref.~\protect\cite{rec}.
       Previous measurements~\protect\cite{said} are
       shown as open circles.} \label{fig:g2}
\end{figure}

Agreement with an existing partial-wave solution
(FA02~\cite{fa02}) was found to be excellent. In fact the FA02
solution was often in better agreement with these new
measurements than those originally included in the fit. The
overall level of agreement with FA02 is illustrated in
Fig.~\ref{fig:g2}.
%%%%%%%%%%%%%%%%%%%%%%%%%%%%%%%%%%%%%%%%%%%%%%%%%%%%%%%%
%%%%%%%%%%%%%%%%%%%%%%%%%%%%%%%%%%%%%%%%%%%%%%%%%%%%%%%%%

%%%%%%%%%%%%%%%%%%%%%%%%%%%%%%%%%%%%%%%%%%%%%%%%
\section{The Reaction $\pi^-p\to\eta n$}

\begin{figure}[htbp]
      \includegraphics[height=0.5\textwidth, angle=90]{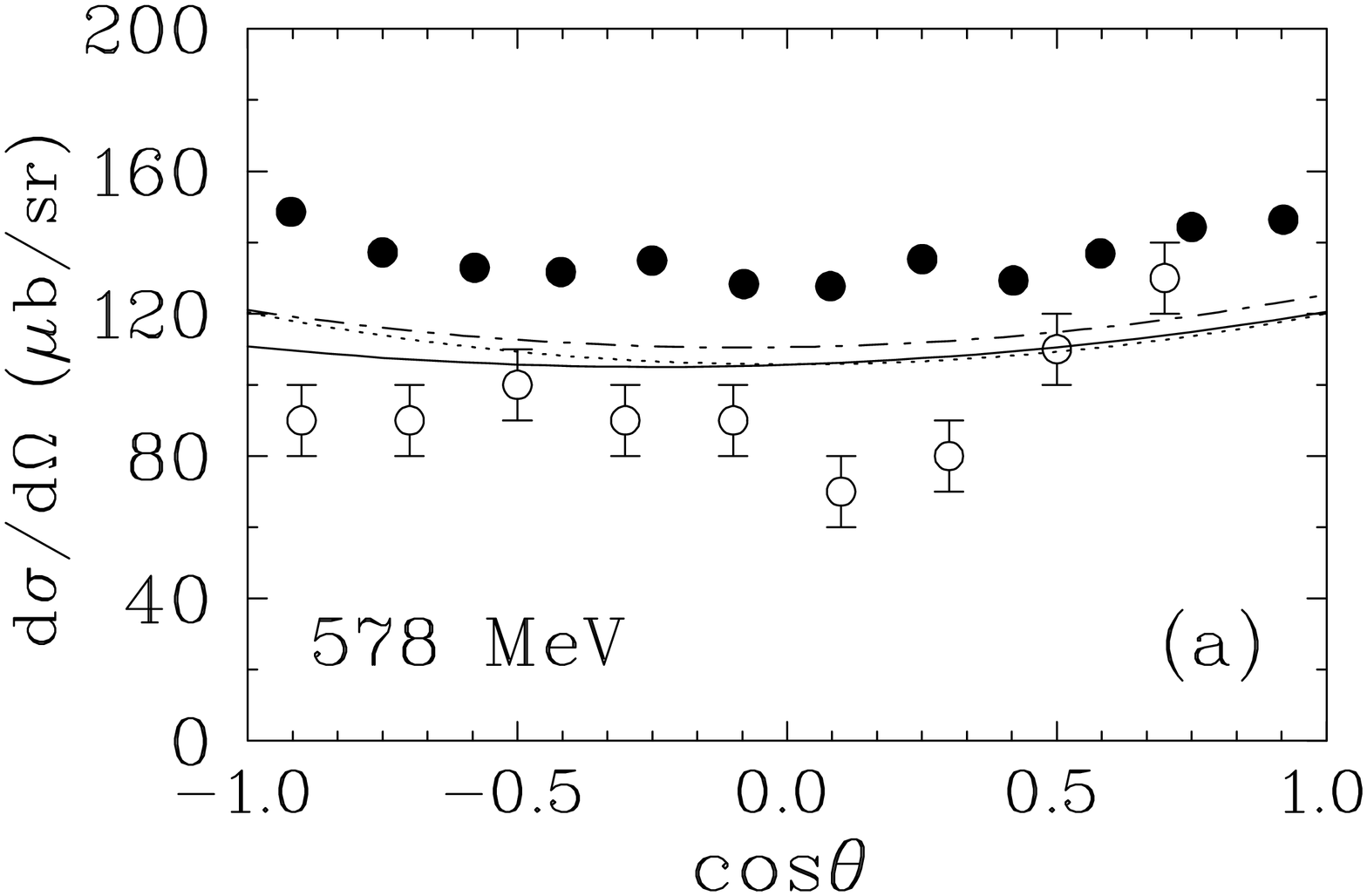}\hfill
      \includegraphics[height=0.5\textwidth, angle=90]{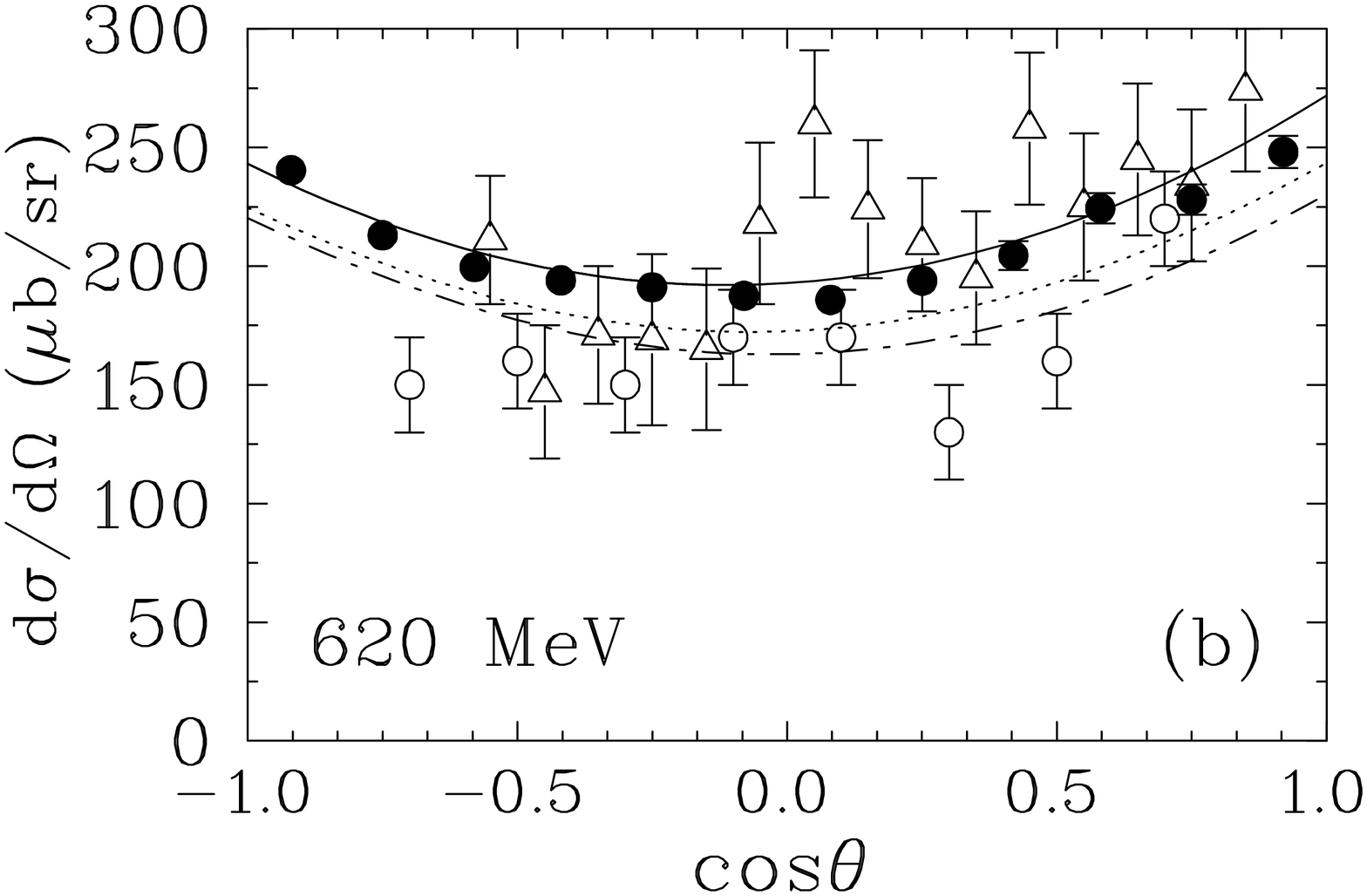}
      \caption{Differential cross sections for
      $\pi^-p\to\eta n$.  FA02~\protect\cite{fa02}
      (E913/E914 data not included), G380, and
      Fit~A~\protect\cite{etan2} shown as solid,
      dash-dotted, and dotted lines, respectively.
      Experimental data are from~\protect\cite{etan1}
      (filled circles), ~\protect\cite{e909} (open
      circles), and ~\protect\cite{said} (open
      triangles) measurements.} \label{fig:g3}
\end{figure}

Differential cross sections for $\eta n$ production were measured
from threshold to T$_{\pi^-}$ = 620~MeV~\cite{etan1} with a goal
to study the $\eta n$ scattering length and $\eta n$ couplings to
the $N(1535)$ and $N(1520)$ resonances. The influence of this
experiment is best illustrated in a plot comparing the Crystal
Ball and prior measurements. In Fig.~\ref{fig:g3} we see how
little the existing data constained single- and multi-channel fits
to this reaction.
%%%%%%%%%%%%%%%%%%%%%%%%%%%%%%%%%%%%%%%%%%%%%%%%%%%%%%%%
%%%%%%%%%%%%%%%%%%%%%%%%%%%%%%%%%%%%%%%%%%%%%%%%%%%%%%%%%

\begin{figure}[htbp]
      \includegraphics[height=0.5\textwidth, angle=90]{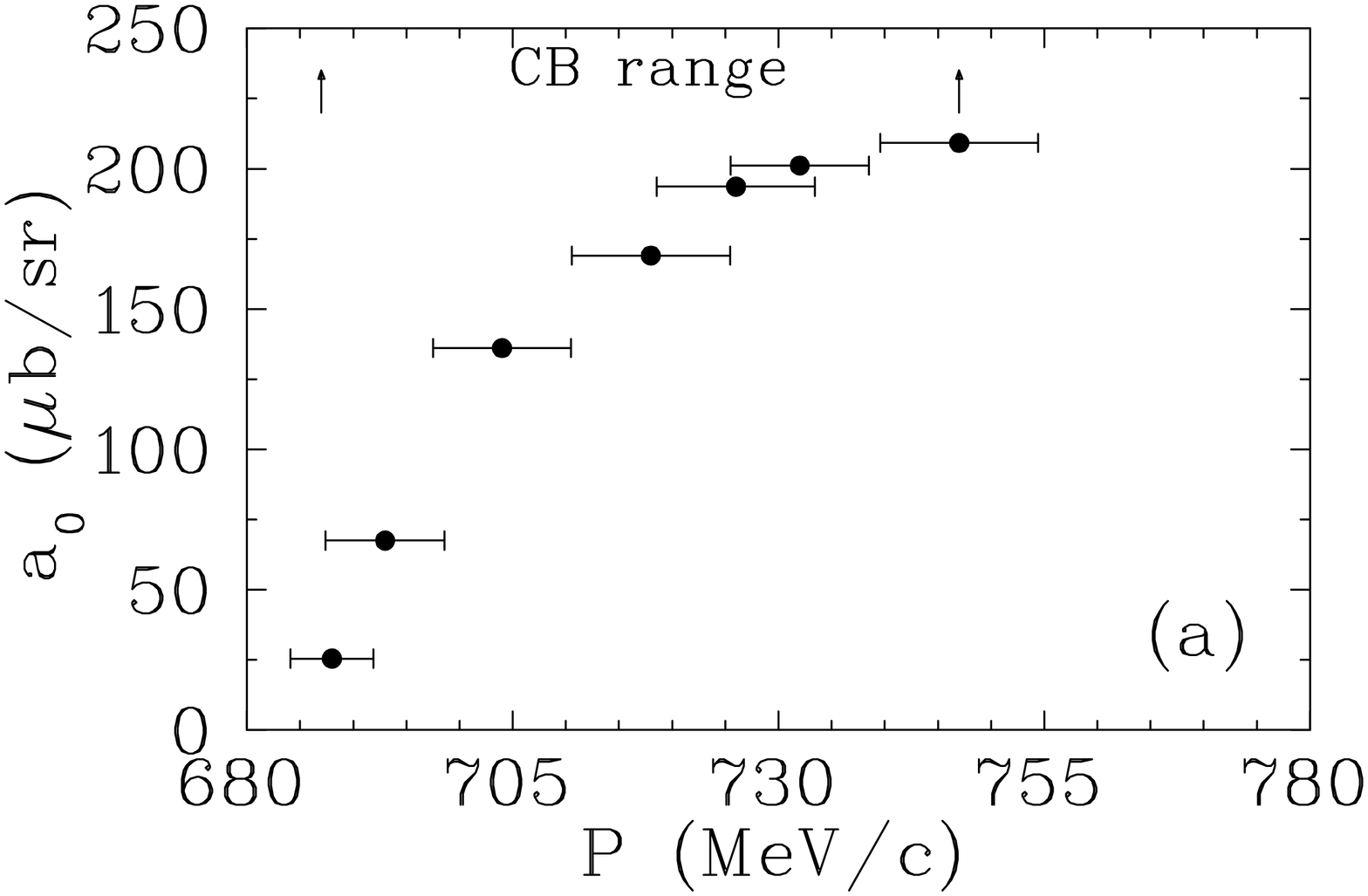}\hfill
      \includegraphics[height=0.5\textwidth, angle=90]{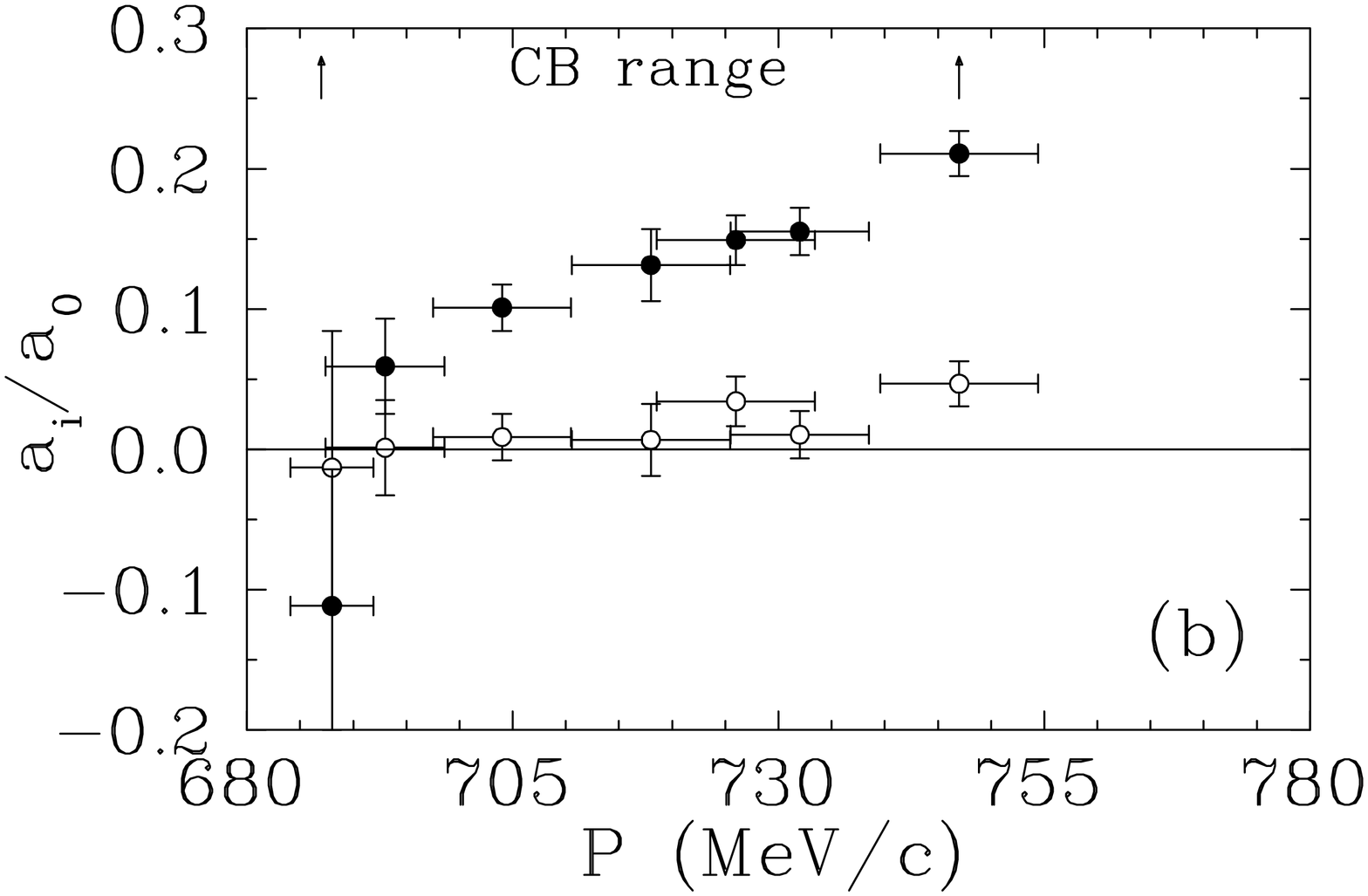}
      \caption{Momentum dependency of (a) the $a_0$
      coefficient of a Legendre polynomial expansion
      of the differential cross sections and (b) ratio
      of the coefficients $a_1/a_0$ (open symbol) and
      $a_2/a_0$ (filled symbol). The horizontal errors
      include the $\pm$2.5~MeV/$c$ beam momentum
      uncertainty and the beam momentum spread of
      E913/E914 data~\protect\cite{etan1}.}
      \label{fig:g4}
\end{figure}

While the dominant contribution is S-wave, due to the $N(1535)$
state, which decays mainly to $\pi N$ and $\eta N$, the Crystal
Ball data clearly reveal a small D-wave component, coming from
the $N(1520)$ resonance.  This is made clear in Fig.~\ref{fig:g4},
where we plot the coefficients of a Legendre polynomial expansion
of the cross sections. Using a K-matrix fit to both the $\pi N$
elastic scattering and $\eta n$ production data, a $N(1520)$
$\eta n$ branching fraction (0.08 to 0.16\%) was
extracted~\cite{etan2}.
%%%%%%%%%%%%%%%%%%%%%%%%%%%%%%%%%%%%%%%%%%%%%%%%%%%%%%%%
%%%%%%%%%%%%%%%%%%%%%%%%%%%%%%%%%%%%%%%%%%%%%%%%%%%%%%%%%

\begin{figure}[htbp]
\centerline{
      \includegraphics[height=0.6\textwidth, angle=90]{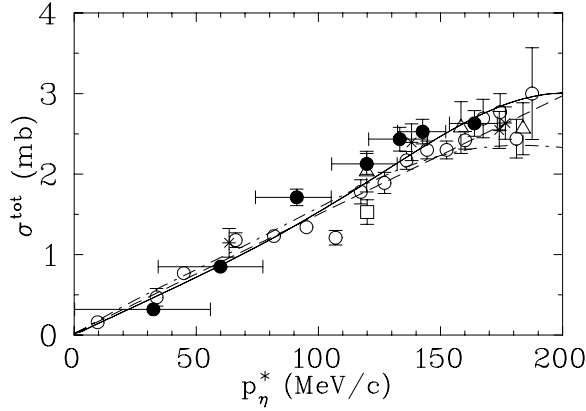}
}
      \caption{$p^{\ast}_{\eta}$ dependence of
      $\sigma^{tot}(\pi^-p\to\eta n)$.  Data and
      notation given in Fig.~\protect\ref{fig:g3}.
      Dashed line shows a linear fit to
      E909~\protect\cite{e909} (open circles) data.}
      \label{fig:g5}
\end{figure}

A plot of $\sigma^{\rm tot} (\pi^-p\to\eta n)$ versus the
eta-meson CM momentum ($p_{\eta}^\ast$) is nearly linear to
almost 200~MeV/c, due to the strong S-wave dominance. This is
illustrated in Fig.~\ref{fig:g5}. Both the Crystal Ball (solid
points) and an earlier BNL measurement~\cite{etan2} (open
circles) are consistent with linearity, though momentum
uncertainty in the Crystal Ball experiment is magnified near
threshold. Using the full near-threshold dataset, we extracted
values for the $\eta n$ scattering length,
(1-1.15)+i(0.3-0.4)~fm, reasonably consistent with the results
of Green and Wycech~\cite{green} and Mosel~\cite{mosel}, if
similar K-matrix methods were used.  Rather different results
followed from the use of a Chew-Mandelstam K-matrix.

%%%%%%%%%%%%%%%%%%%%%%%%%%%%%%%%%%%%%%%%%%%%%%%%%%%%%%%%
%%%%%%%%%%%%%%%%%%%%%%%%%%%%%%%%%%%%%%%%%%%%%%%%%%%%%%%%%
\begin{figure}[htbp]
      \includegraphics[height=0.5\textwidth, angle=90]{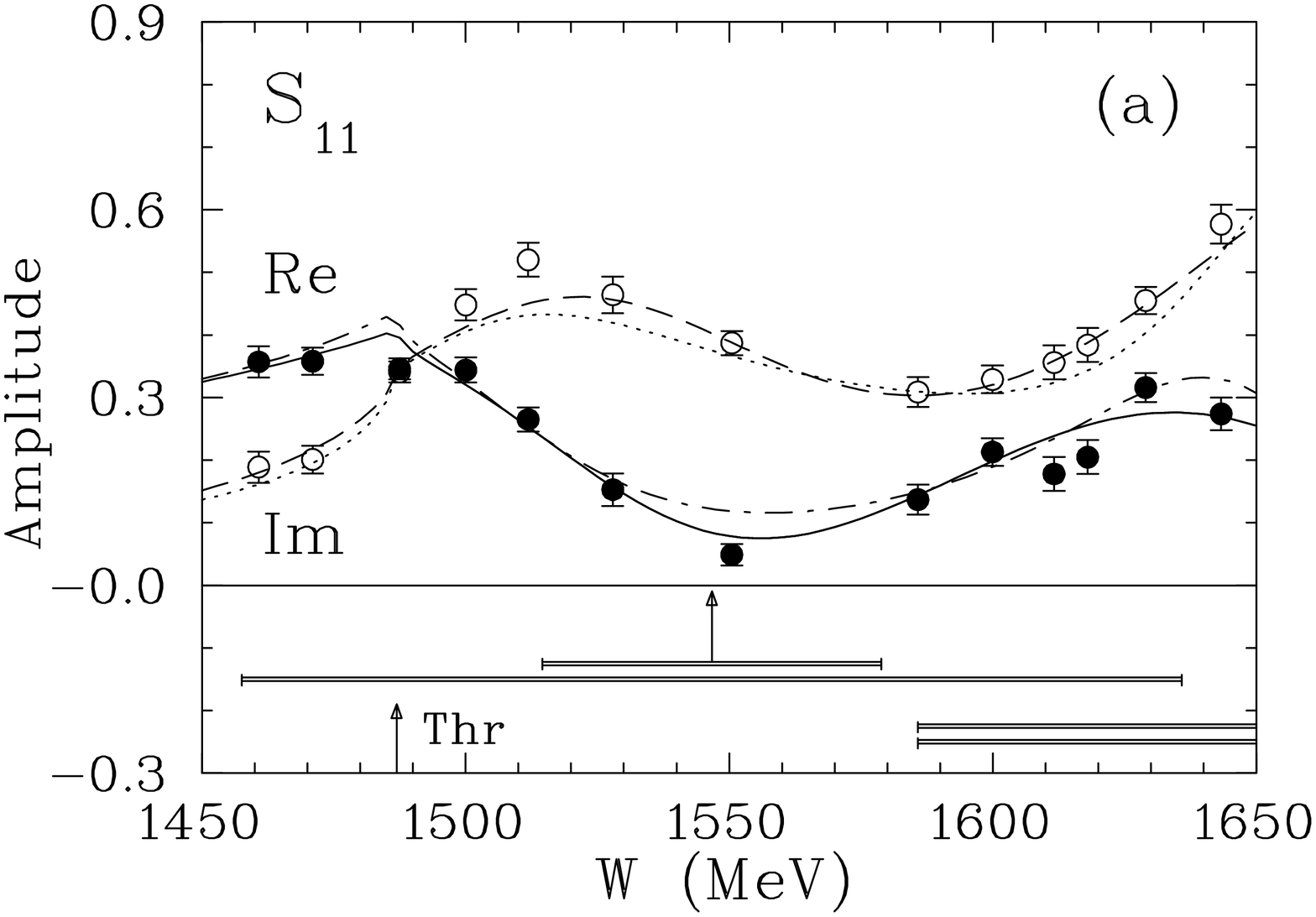}\hfill
      \includegraphics[height=0.5\textwidth, angle=90]{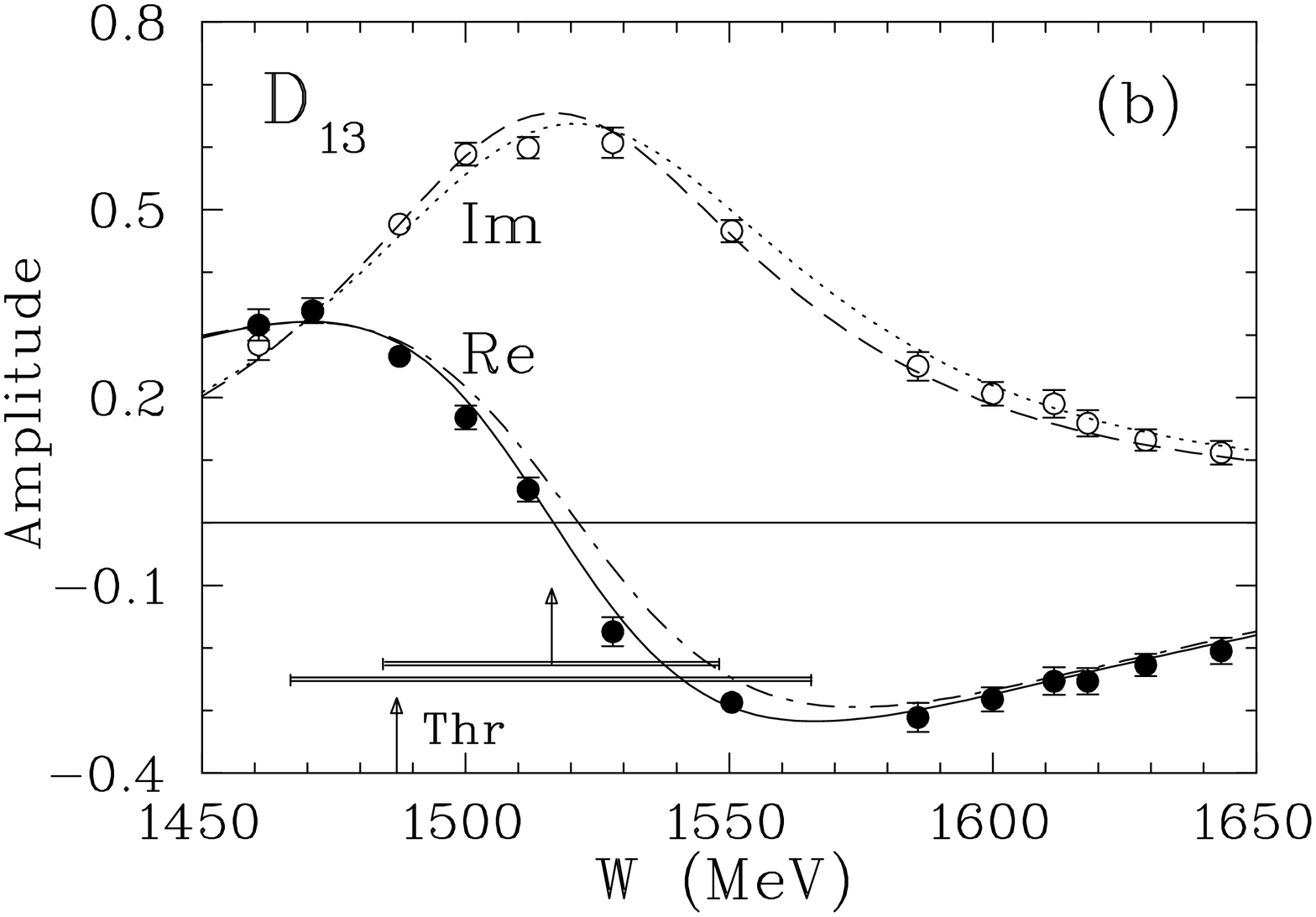}
      \caption{(a) $S_{11}$ and (b) $D_{13}$ partial
      amplitudes for $\pi N$ elastic scattering.  Solid
      (dashed) curves give the real (imaginary) parts of
      amplitudes corresponding to the predictions of
      solution FA02~\protect\cite{fa02} (E913/E914 data
      not included). Single-energy solutions associated
      with FA02 are plotted as filled and open circles.
      Dash-dotted (dotted) curves show the real (imaginary)
      parts of amplitudes corresponding to G380.
      Differences between G380 and Fit~A are not
      significant.  All amplitudes are dimensionless.
      Vertical arrows indicate $W_R$ and horizontal bars
      show full $\Gamma$/2 and partial widths for
      $\Gamma_{\pi N}$ associated with the FA02 results.}
      \label{fig:g6}
\end{figure}
%%%%%%%%%%%%%%%%%%%%%%%%%%%%%%%%%%%%%%%%%%%%%%%%%%%%%%%%%
\begin{figure}[htbp]
      \includegraphics[height=0.5\textwidth, angle=90]{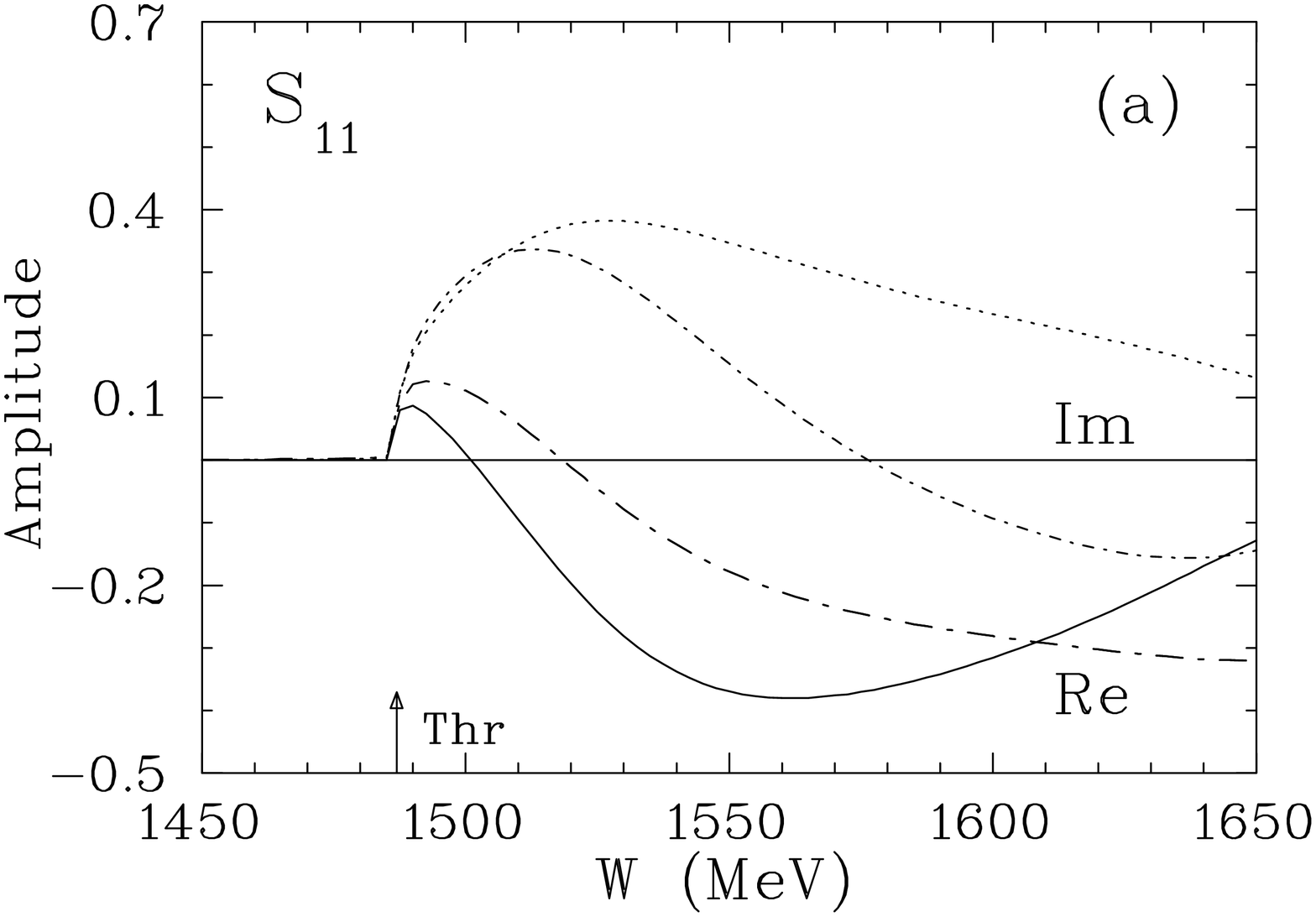}\hfill
      \includegraphics[height=0.5\textwidth, angle=90]{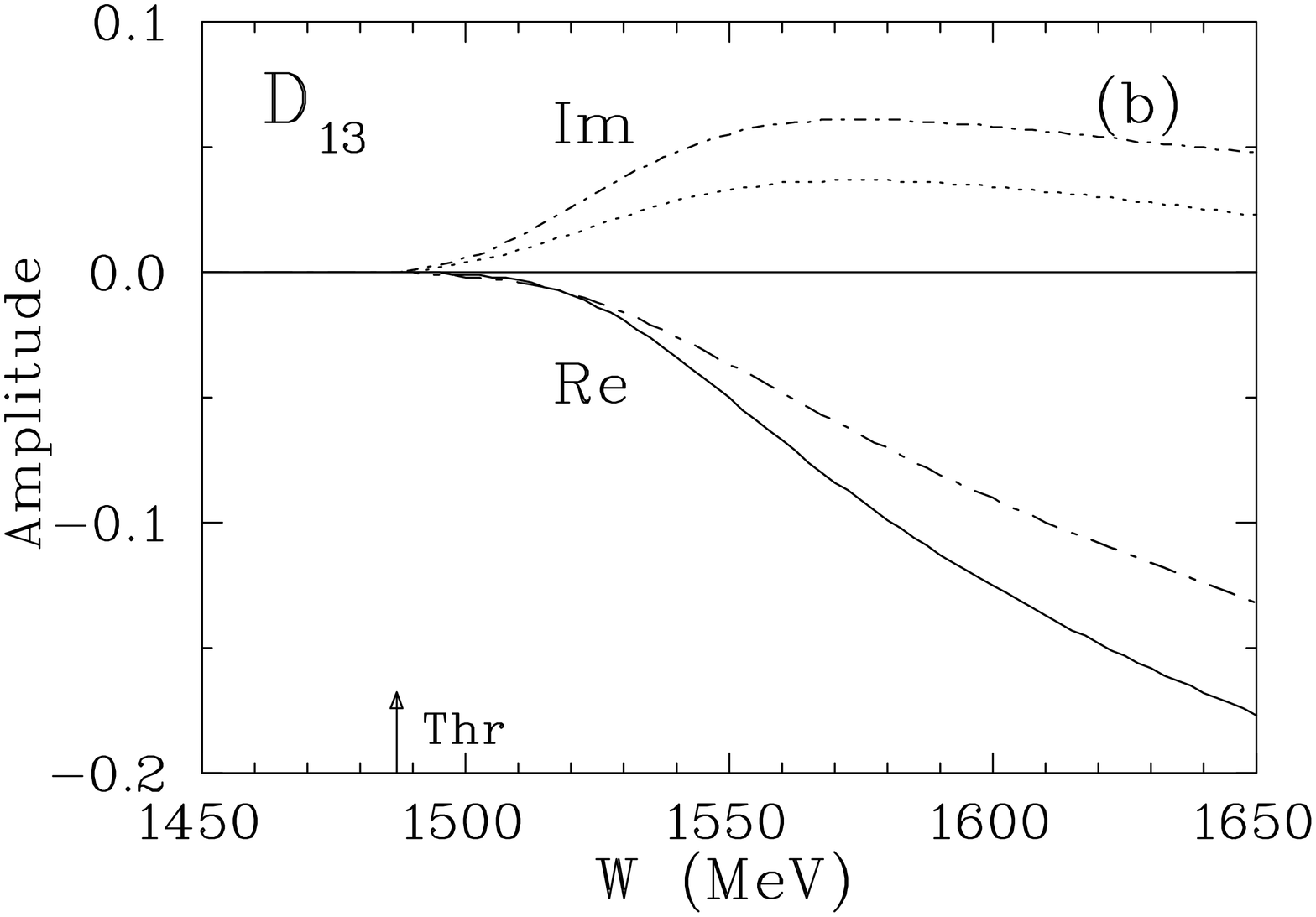}
      \caption{(a) $S_{11}$ and (b) $D_{13}$ partial amplitude for
          $\pi^-p\to\eta n$.  Dash-dotted (dotted) curves show
          the real (imaginary) parts of amplitudes corresponding
          to G380.  Solid (short-dash-dotted) lines represent
          the real (imaginary) parts of amplitudes corresponding
          to the Fit~A.} \label{fig:g7}
\end{figure}
%%%%%%%%%%%%%%%%%%%%%%%%%%%%%%%%%%%%%%%%%%%%%%%%%%%%%%%%%
\begin{figure}[htbp]
\centerline{
      \includegraphics[height=0.5\textwidth, angle=90]{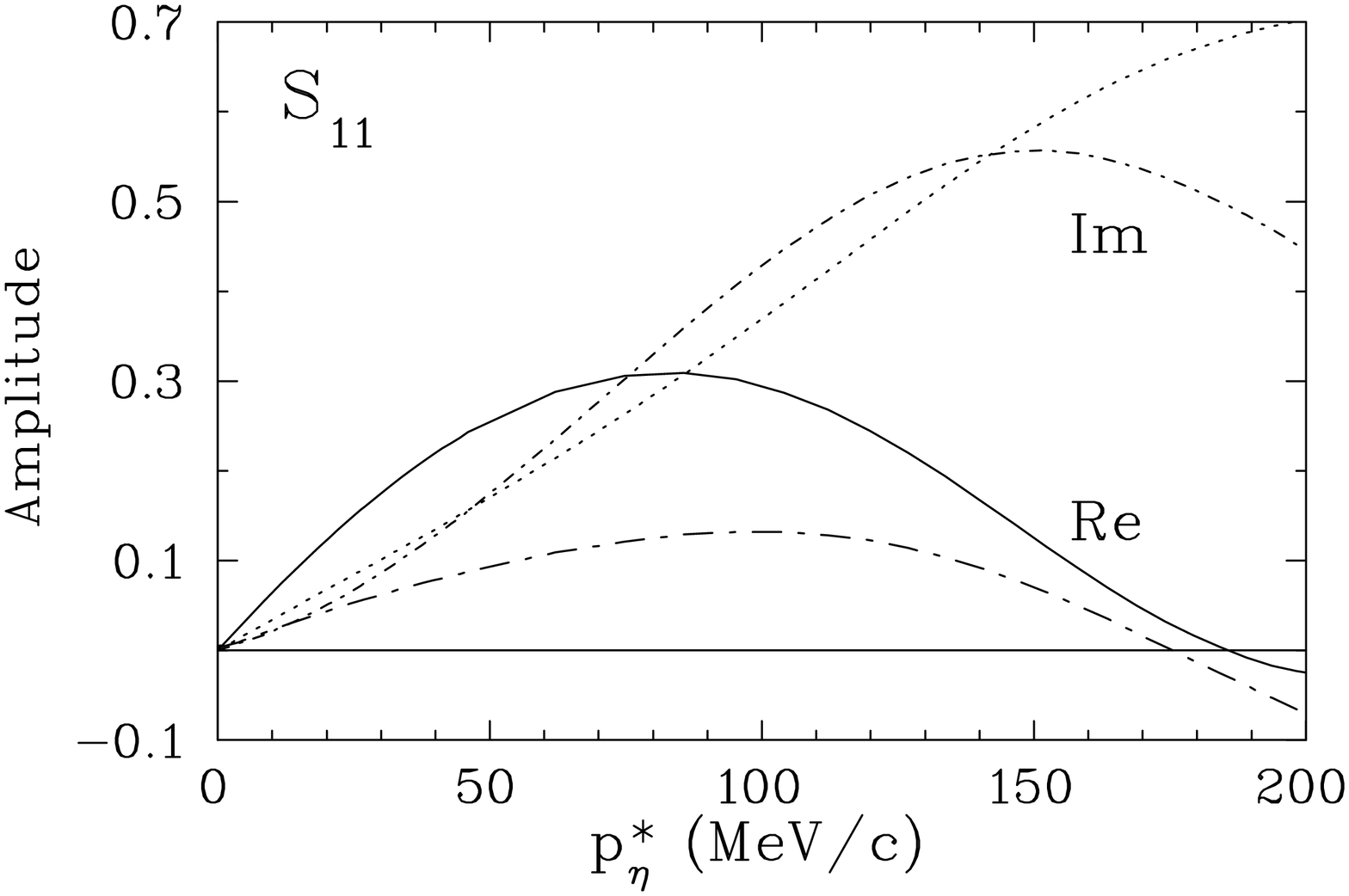}
}
      \caption{$p^{\ast}_{\eta}$ dependence of the $S_{11}$
          amplitude for the reaction $\eta n\to\eta n$.
          Dash-dotted (dotted) curves give the real (imaginary)
          parts of amplitudes corresponding to the solution
          G380.  Solid (short-dash-dotted) lines represent the
          real (imaginary) parts of amplitudes Fit~A.}
          \label{fig:g8}
\end{figure}

Amplitude results using the Chew-Mandelstam K-matrix (G380) versus
a K-matrix approach similar to that used in Ref.~\cite{green}
(Fit~A) are compared in Figs.~\ref{fig:g6}--~\ref{fig:g8}.  For
the most important $\pi N$ amplitudes, $S_{11}$ and $D_{13}$, the
differences are minor. Much larger deviations are seen in the
amplitudes for $\pi N\to \eta N$ and $\eta N\to \eta N$.

New results from a comprehensive partial-wave analysis of
$\pi^\pm p$ elastic scattering and charge-exchange data, covering
the region from threshold to 2.6~GeV in the lab pion kinetic
energy, employed a coupled-channel formalism to simultaneously
fit $\pi^-p\to\eta n$ data to 0.8~GeV~\cite{sp06}.  This fit
included all of the E913/E914 data mentioned above.  Breit-Wigner
parameters for resonances are listed in Table~\ref{tbl}. In the
SP06 and FA02 fits, a unitary Breit-Wigner plus background form
was assumed for the resonant partial wave.  Data within an energy
bin were then fitted using this representation.  The remaining
waves were fixed to values found in the full global (SP06)
analysis.

%%%%%%%%%%%%%%%%%%%%%%%%%%%%%%%%%%%%%%%%%%%%%%%%%%%%%%%
\begin{table}[th]
\tbl{Resonance couplings from a Breit-Wigner fit
         to the recent SP06~\protect\cite{sp06}, our
         previous FA02~\protect\cite{fa02} solutions,
         and a range from the {[}RPP{]}~\protect\cite{rpp}
         (in square brackets).  Masses W$_R$, widths
         $\Gamma$, and partial width $\Gamma_{\pi
         N}$/$\Gamma$ are listed for isospin $1/2$
         baryon resonances relevant to the Crystal Ball
         measurements.}
{\footnotesize
\begin{tabular}{ccccc}
\colrule
Resonance      & W$_R$ & $\Gamma$    & $\Gamma_{\pi N}/ \Gamma$ & Ref\\
               & (MeV) & (MeV)       &                          &    \\
\colrule
N(1440)$P_{11}$& 1485.0$\pm$1.2&284$\pm$18    &0.787$\pm$0.016    & SP06\\
               & 1468.0$\pm$4.5&360$\pm$26    &0.750$\pm$0.024    & FA02\\
               &[1420$-$1470]  &[200$-$450]   &[0.55$-$0.75]      & RPP\\
N(1520)$D_{13}$& 1514.5$\pm$0.2&103.6$\pm$0.4 &0.632$\pm$0.001    & SP06\\
               & 1516.3$\pm$0.8&98.6$\pm$2.6  &0.640$\pm$0.005    & FA02\\
               &[1515$-$1525]  &[100$-$125]   &[0.55$-$0.65]      & RPP\\
N(1535)$S_{11}$& 1547.0$\pm$0.7&188.4$\pm$3.8 &0.355$\pm$0.002    & SP06\\
               & 1546.7$\pm$2.2&178.0$\pm$11.6&0.360$\pm$0.009    & FA02\\
               &[1525$-$1545]  &[125$-$175]   &[0.35$-$0.55]      & RPP\\
\colrule
\end{tabular}\label{tbl}}
\end{table}

%%%%%%%%%%%%%%%%%%%%%%%%%%%%%%%%%%%%%%%%%%%%%%%%%%%%%%%%%

%%%%%%%%%%%%%%%%%%%%%%%%%%%%%%%%%%%%%%%%%%%%%%%%
\section{Inverse Pion Photoproduction}

The inverse pion-photoproduction reaction ($\pi^- p\to \gamma n$)
has been measured at 18 values of T$_{\pi^-}$ from 136 to
621~MeV~\cite{rec}.  Using detailed balance, these yield angular
distributions for the photoproduction reaction ($\gamma n\to\pi^-p$)
for values of $E_{\gamma}$ between 285 and 769~MeV, spanning the
$N(1440)$ resonance region. This procedure avoids the complication
of extracting neutron-target information from the deuteron in a
direct measurement of $\pi^-$ photoproduction.

\begin{figure}[htbp]
      \includegraphics[height=0.5\textwidth, angle=90]{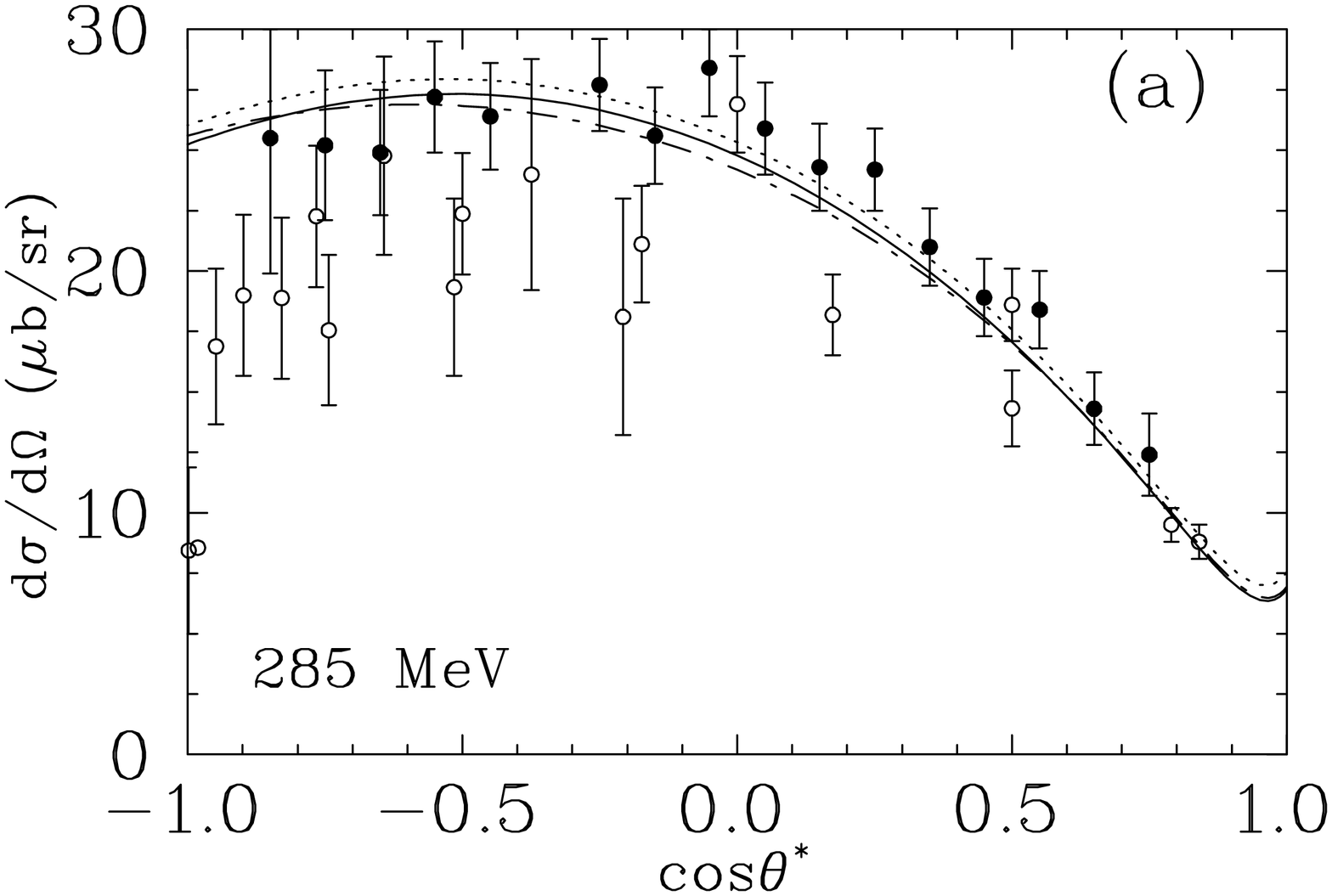}\hfill
      \includegraphics[height=0.5\textwidth, angle=90]{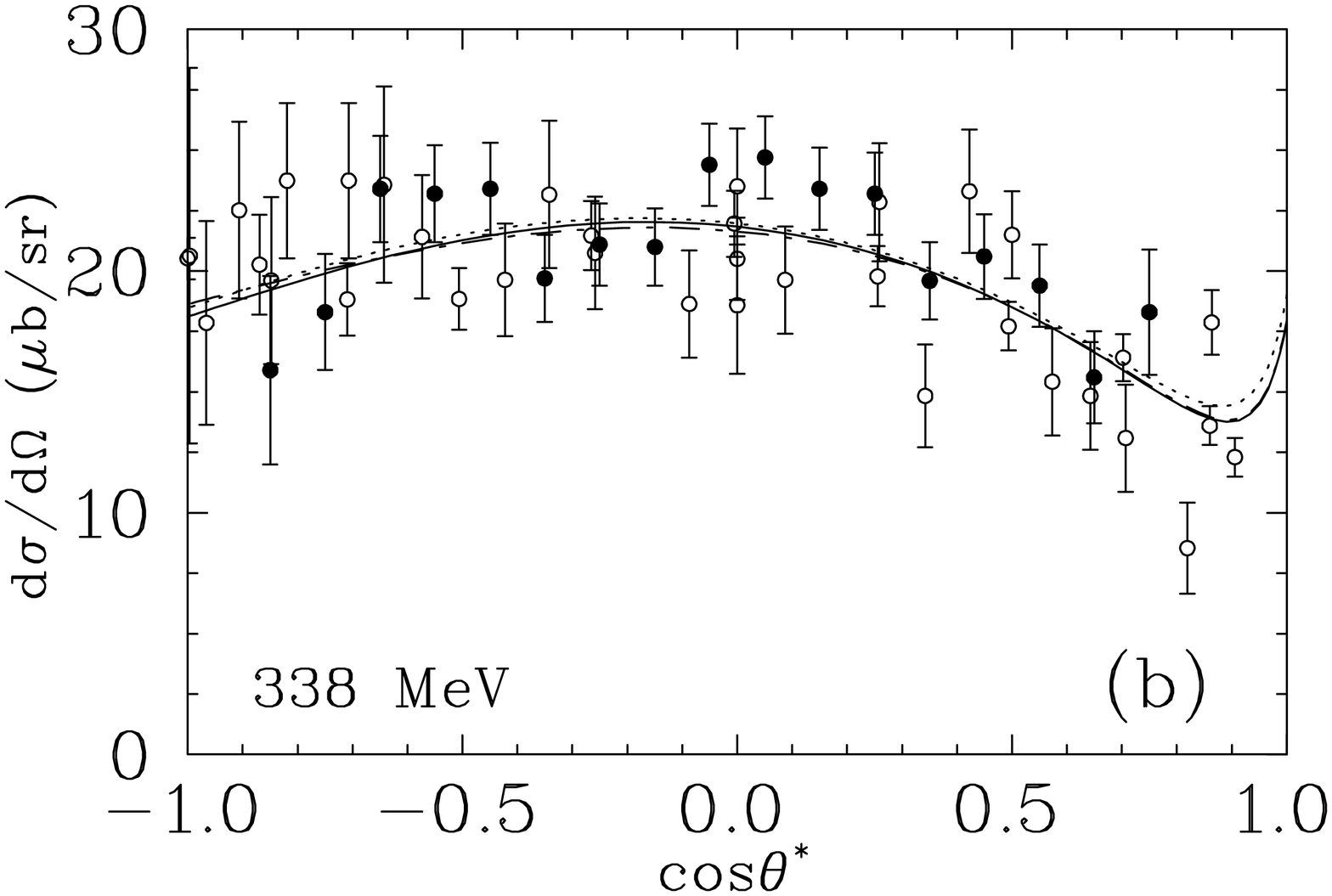}
      \caption{Differential cross sections for $\gamma
      n\to\pi^-p$.  Dashed-dotted (solid) curves correspond
      to SM02 (SH04) solution~\protect\cite{sm02}.  The MAID
      solution~\protect\cite{maid} are plotted with dashed
      lines.  Experimental data are from
      E913/E914~\protect\cite{rec} (filled circles)
      and~\protect\cite{said} (open circles) measurements.}
      \label{fig:g9}
\end{figure}

The Crystal Ball experiment has effectively doubled the $\pi^-$
photoproduction database over the $N(1440)$ region, and has been
included in a modified multipole analysis. These new data were
found to agree very well with predictions from the SAID and MAID
multipole solutions. Only slight modifications of the SAID
multipoles were required to obtain an overall $\chi^2$ near
unity for these data. A comparison with SAID and MAID results is
interesting. In some cases, the predictions follow the Crystal
Ball data while missing older measurements that were included in
the SAID and MAID fits. An example is given in Fig.~\ref{fig:g9}.
%%%%%%%%%%%%%%%%%%%%%%%%%%%%%%%%%%%%%%%%%%%%%%%%%%%%%%%%
%%%%%%%%%%%%%%%%%%%%%%%%%%%%%%%%%%%%%%%%%%%%%%%%%%%%%%%%%
\begin{figure}[htbp]
      \includegraphics[height=0.5\textwidth, angle=-90]{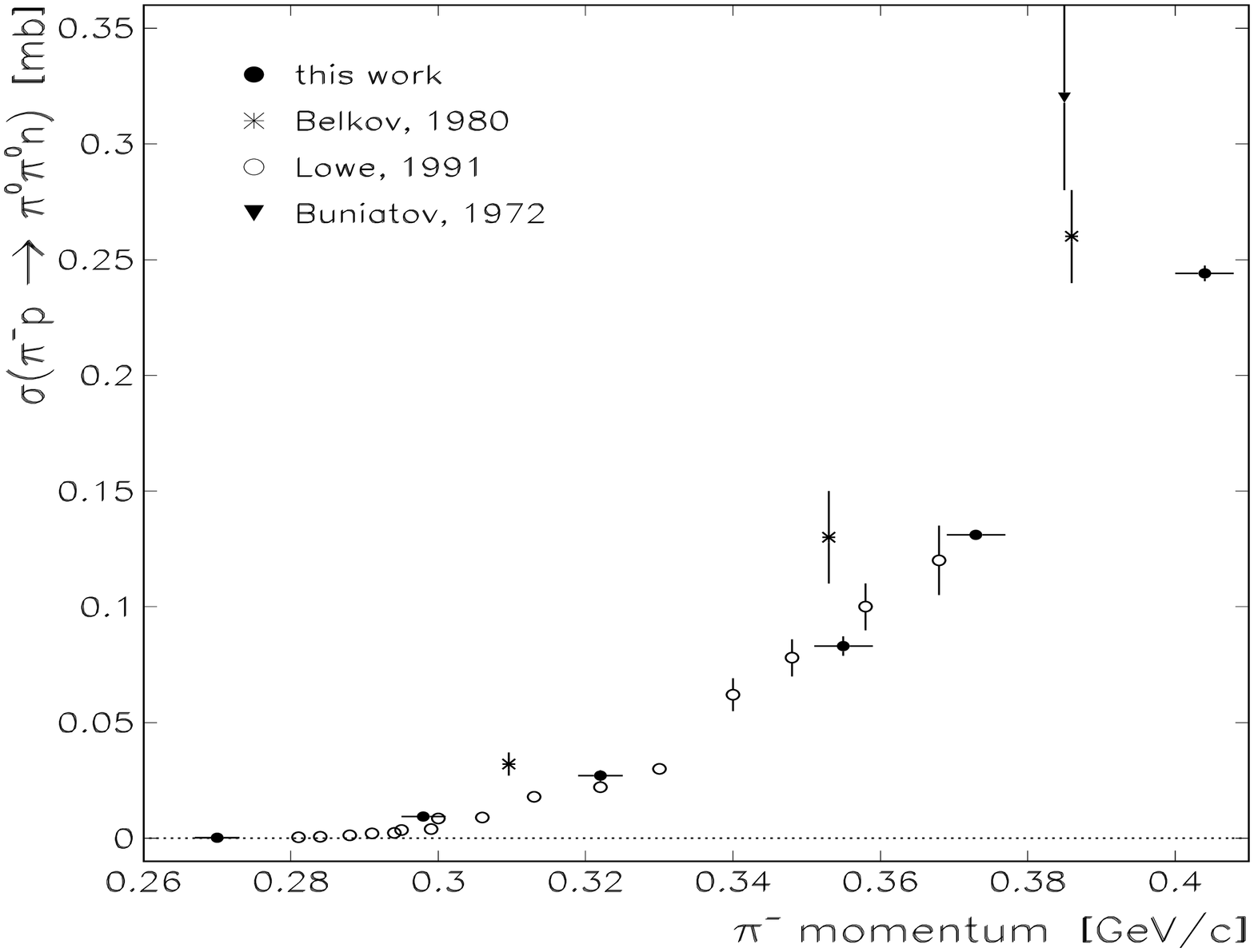}\hfill
      \includegraphics[height=0.5\textwidth, angle=-90]{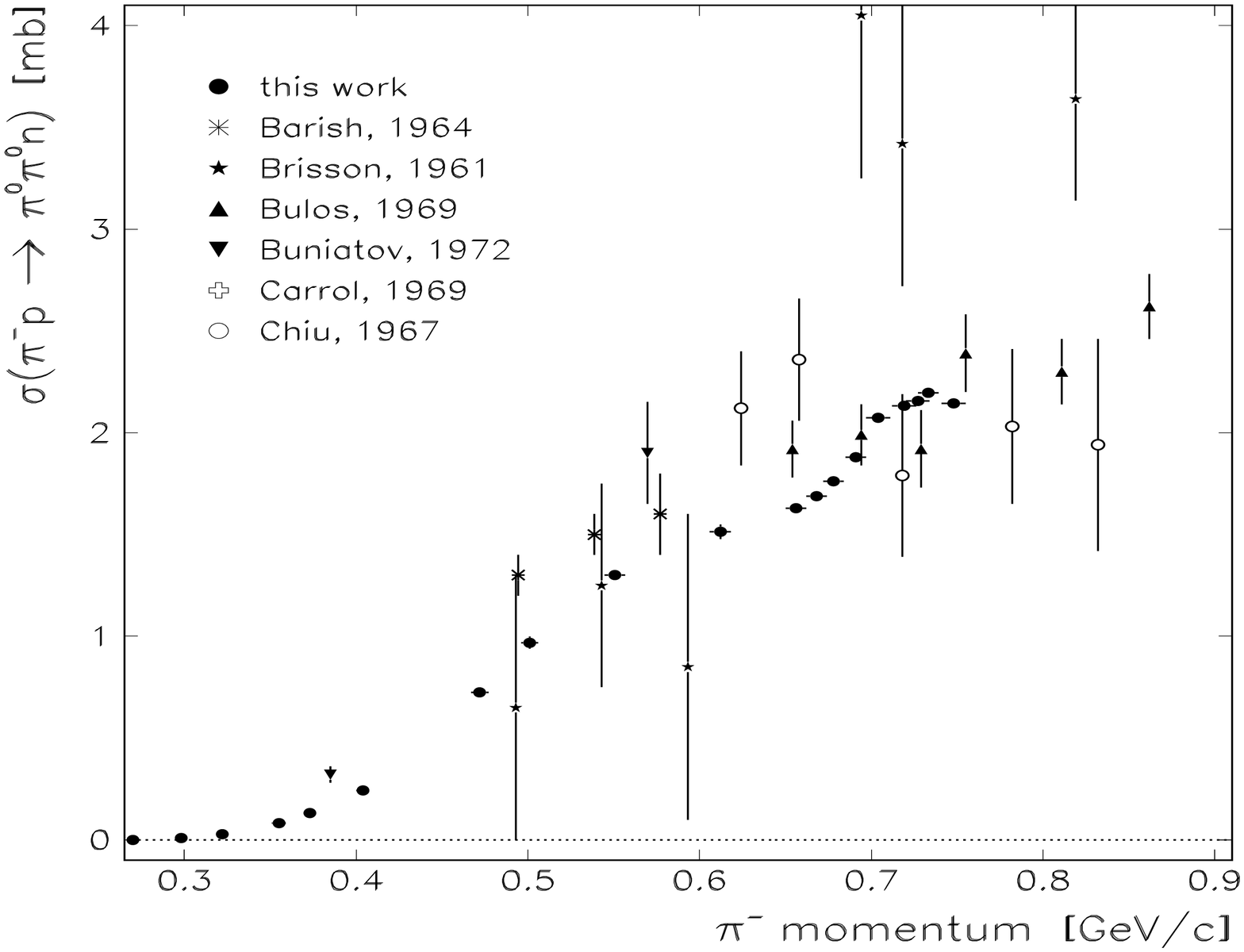}
      \caption{Total cross sections for the
      $\pi^-p\to\pi^0\pi^0n$ reaction.  Experimental
      data are from E913/E914~\protect\cite{pi2pi}
      (filled circles) and previous~\protect\cite{pi2pi}
      measurements.} \label{fig:g10}
\end{figure}
%%%%%%%%%%%%%%%%%%%%%%%%%%%%%%%%%%%%%%%%%%%%%%%%%%%%%%%%%

%%%%%%%%%%%%%%%%%%%%%%%%%%%%%%%%%%%%%%%%%%%%%%%%%%%%%%%%
\section{The Reaction $\pi^-p\to\pi^0\pi^0n$}

Measurements of the reaction $\pi^-p\to\pi^0\pi^0n$ were
made from 160~MeV, just above threshold, to
625~MeV~\cite{pi2pi}.  The resulting total cross sections
confirmed the trend found in an earlier near-threshold
experiment~\cite{Lowe}.  At higher energies, the Crystal
Ball results are far more precise than any previous
determination.  These two kinematic ranges are displayed in
Fig.~\ref{fig:g10}.

Dalitz plots of the 3-body final state suggested the
dominance of a $\pi^0\Delta^0$ intermediate state, formed
through the sequential process
\begin{equation}
\pi^-p\to N^\ast\to\pi^0\Delta^0(1232)\to\pi^0\pi^0n.
\end{equation}
The contribution of another sequential process
\begin{equation}
\pi^-p\to N^\ast\to f_0(600)n\to\pi^0\pi^0n
\end{equation}
was also considered. (The $f_0(600)$ is a PDG designation for
the more commonly known $\sigma$ meson.) No direct evidence
was found for this contribution, suggesting the need for a more
detailed partial-wave analysis.

A number of recent calculations~\cite{theory} have incorporated
these measurements. Approaches explicitly including resonance
contributions (the HB$\chi$PT calculation did not), found
important contributions from both of the above processes, with
the most significant $N^\ast$ being the $N(1440)$ state.

\section{The Future - the Crystal Ball and TAPS at MAMI}

\begin{figure}[htb]
\centerline{ \psfig{file=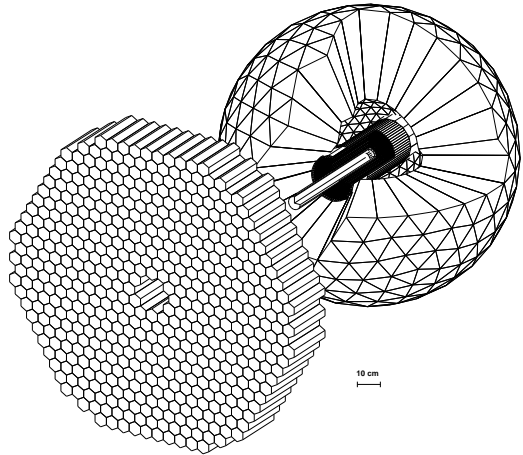,width=0.6\textwidth}} \caption{The
experimental apparatus for our measurements at MAMI. To show the
position of the cylindrical wire chamber and the target inside the
Crystal Ball some of the Crystal Ball crystals have been omitted.}
\label{fig:apparatus}
\end{figure}

A program of photonuclear physics is in progress using the
combination of the $4\pi$ Crystal Ball and TAPS
(Fig.~\ref{fig:apparatus}) at MAMI. The facilities at MAMI include
tagged, polarized photons and polarized $p$, $d$, and $^3He$
targets. Such a high-powered facility allows us to measure meson
photoproduction and a variety of double polarization observables.
The ``Crystal Ball at MAMI" is now a top priority of the
laboratory.

The combination of tagged photons and the Crystal Ball and TAPS
provides an excellent opportunity for outstanding physics. MAMI B
produces photons up to 800 MeV in 2-MeV bins. The MAMI C will
provide photons up to 1400 MeV. The TAPS detector acting as the
forward detection array of the CB, substantially improving the
acceptance at forward angles. This provides almost-4$\pi$ coverage
and a unique opportunity to measure reactions in which one or more
of the final-state particles decay into photons. The Crystal Ball
and TAPS also have a high detection efficiency for neutrons. This,
coupled with the excellent energy resolution of the MAMI tagged
photon beam, provides an opportunity to make precision
measurements of the heretofore sparsely-measured neutral
photoproduction channels.

%%%%%%%%%%%%%%%%%%%%%%%%%%%%%%%%%%%%%%%%%%%%%%%%%%
\section*{Acknowledgments}

This work as supported in part by the U.S. Department of
Energy grant DE-FG02-99ER41110 and by funding from Jefferson
Lab.

%%%%%%%%%%%%%%%%%%%%%%%%%%%%%%%%%%%%%%%%%%%%%%%%%%%%%%%

%%%%%%%%%%%%%%%%%%%%%%%%%%%%%%%%%%%%%%%%%%%%%5

\begin{thebibliography}{0}
\bibitem{cex1} M.E.~Sadler \textit{et al.} [Crystal Ball
               Collaboration], \textit{Phys. Rev. C}
               \textbf{69}, 055206 (2004).
\bibitem{cex2} A.~Starostin \textit{et al.} [Crystal Ball
               Collaboration], \textit{Phys. Rev. C}
               \textbf{72}, 015205 (2005).
\bibitem{fa02} R.A.~Arndt, W.J.~Briscoe, I.I.~Strakovsky,
               R.L.~Workman, and M.M.~Pavan, \textit{Phys.
               Rev. C} \textbf{69}, 035213 (2004);
               SAID solutions at http://gwdac.phys.gwu.edu.
\bibitem{rec}  A.~Shafi \textit{et al.} [Crystal Ball
               Collaboration], \textit{Phys. Rev. C}
               \textbf{70}, 035204 (2004).
\bibitem{said} The full database and numerous PWAs can be
               accessed via an ssh call to the SAID
               facility \hbox{gwdac.phys.gwu.edu}, with
               userid: said (no password), or a link to
               the website \hbox{http://gwdac.phys.gwu.edu}.
\bibitem{etan1}S.~Prakov \textit{et al.} [Crystal Ball
               Collaboration], \textit{Phys. Rev. C}
               \textbf{72}, 015203 (2005).
\bibitem{e909} T.~W.~Morrison \textit{et al.}, \textit{Bull.
               Am. Phys. Soc.} \textbf{45}, 58 (2000);
               T.W.~Morrison, Ph.~D. Thesis, The George
               Washington University, Dec.~1999.
\bibitem{etan2}R.A.~Arndt, W.J.~Briscoe, T.W.~Morrison,
               I.I.~Strakovsky, R.L.~Workman, and A.B.~Gridnev
               \textit{Phys. Rev. C} \textbf{72}, 045202 (2005).
\bibitem{green}A.M.~Green and S.~Wycech, \textit{Phys. Rev. C}
               \textbf{71}, 014001 (2005).
\bibitem{mosel}G.~Penner and U.~Mosel, \textit{Phys. Rev. C}
               \textbf{66}, 055211 (2002).
\bibitem{sm02} R.A.~Arndt, W.J.~Briscoe, I.I.~Strakovsky,
               and R.L.~Workman, \textit{Phys. Rev. C}
               \textbf{66}, 055213 (2002).
\bibitem{maid} D.~Drechsel, O.~Hanstein, S.S.~Kamalov, and
               L.~Tiator, \textit{Nucl. Phys.} \textbf{A645},
               45 (1999).
\bibitem{pi2pi}S.~Prakov \textit{et al.} [Crystal Ball
               Collaboration], \textit{Phys. Rev. C}
               \textbf{69}, 045202 (2005).
\bibitem{Lowe} J.~Lowe \textit{et al.} \textit{Phys. Rev. C}
               \textbf{44}, 956 (1991).
\bibitem{theory}N.~Mobed, J.~Zhang, and D.~Singh,
               \textit{Phys. Rev. C} \textbf{72}, 045204 (2005);
               H.~Kamano and M.~Arima,
               \textit{Phys. Rev. C} \textbf{73}, 055203 (2006);
               S.~Schneider, S.~Krewald, and Ulf-G.~Meissner,
               nucl-th/06030.
\bibitem{sp06} R.A.~Arndt, W.J.~Briscoe, I.I.~Strakovsky, and
               R.L. Workman, `Extended partial-wave analysis of
               $\pi N$ scattering data,' submitted to
               \textit{Phys. Rev. C} [nucl--th/0605082].
\bibitem{rpp}  S.~Eidelman \textit{et~al.}, \textit{Review of
               Particle Physics}, \textit{Phys. Lett. B}
               \textbf{592}, 1 (2004).
\end{thebibliography}
\end{document}